\documentclass[twocolumn,aps,showpacs,amsmath,preprintnumbers,floatfix,a4paper]{revtex4}

\usepackage{graphicx}
\usepackage{dcolumn}
\usepackage{bm}
\usepackage{color}
\usepackage{amssymb}
\usepackage{natbib}

\begin{document}
\title{Coherent Transition Radiation from Relativistic Beam-Foil Interaction in the Terahertz and Optical Range}

\author{W. J. Ding$^{1}$} 
	\thanks{dingwj@ihpc.a-star.edu.sg}
\author{F. Y. Li$^{2}$}
\author{S. M. Weng$^{3,4}$}
\author{P. Bai$^{1}$}
\author{Z. M. Sheng$^{2,3,4}$}

\affiliation
{
	$^1$ Institute of High Performance Computing, Agency for Sicence Tenchnology and Research, Singapore 138632\\
	$^2$ SUPA, Department of Physics, University of Strathclyde, Glasgow G4 0NG, UK\\
	$^3$ Key Laboratory for Laser Plasmas (MoE) and Department of Physics and Astronomy, Shanghai Jiao Tong University, Shanghai 200240, China\\
	$^4$ Collaborative Innovation Center of IFSA, Shanghai Jiao Tong University, Shanghai 200240, China
}

\date{\today}

\begin{abstract}
	Coherent transition radiation (CTR) from relativistic electron beam interaction with an overdense plasma foil is investigated by making use of two-dimensional particle-in-cell simulations. Well-defined single electron beam either of uniform profile or having substructures is considered for various beam-plasma parameters. The main purpose is to mimic the complicated beam-plasma conditions that is often found, for example, in intense laser plasma interactions. Key properties of the CTR concerning their temporal, angular and spectral profiles are identified. Several saturation effects due to the beam energy, size and foil density are found for the CTR energy, and the dependences vary for different spectral components such as in the Terahertz (THz) and optical range. The detailed substructure of the beam also affects greatly the radiation generation, leading to distinctive high harmonic components. Electrons with kinetic energy from sub MeV to tens of GeV are explored. For few MeV electron beams, the effects of the foil plasma on the beam dynamics and associated CTR generation, resembles closely the CTR from hot electrons produced in intense laser-plasma interactions. These results may find important applications in beam diagnostics either in laser-plasma based acceleration or conventional accelerators. They may also be employed to design novel THz radiation sources using tunable electron beams. 
\end{abstract}


\maketitle

\section{Introduction}
Transition radiation (TR) refers to the electromagnetic radiation emitted when charged particles traverse the boundary between two different media. It was first brought forward by Ginzburg and Frank more than half a century ago~\cite{ginzburg1946radiation}. Of particular interest is the coherent transition radiation (CTR) which occurs for certain spectral ranges given the charged particles are somewhat bunched in space time. Dependent on the charged beam energy and size, the CTR or TR has attracted intensive studies in a broad spectral range covering from X-ray, Terahertz (THz)~\cite{ding2013high,ding2016sub,liao2016demonstration, Mondal2017nano} to microwaves~\cite{wu2016relativistic}.

In this paper, we investigate the transition radiation, especially coherent transition radiation (CTR), emitted by an electron beam interaction with an overdense plasma or metal foil. The goal of our study is to extract the characteristic features which are most important for experimental work. This is initiated from the study of CTR from laser-plasma interaction, where the charateristics of electron beam is rather complicated. While here the parameters of electron beam are manually controlled so a clear effect on the radiation is found. This could be a guidance for the future experiments of CTR from laser-plasma interaction.  
A theoretical model of CTR by electron beam is given in Sec. II. Numerical simulations via particle-in-cell (PIC) codes are performed. Results of CTR from a uniform electron beam are demonstrated in Sec. III. Electron beam with sub-structures is considered in Sec. IV. This is considered because in laser-plasma interaction, continual electron bunches are usually produced. A summary is given in Sec. V.

\section{Analytical angular spectrum of  CTR by an electron beam}
In the incidence plane, the energy spectrum (per frequency interval per solid angle) of the TR by a single electron passing through a target boundary is given by~\cite{zheng2003theoretical}:
\begin{equation}
	\label{spec_single}
\frac{d^2\xi}{d\omega d\Omega} = \frac{e^2}{\pi^2 c} |S(\beta, \varphi, \phi)|^2,
\end{equation}
where $e$ is the elementary  charge, $c$ the light velocity in free space, $\beta$ the electron velocity normalized to $c$, $\varphi$ the electron injection direction relative to the target normal, $\phi$ the observation direction, and  
\begin{equation}
S(\beta, \varphi, \phi) = \frac{\beta\cos\varphi (\sin\phi-\beta\sin\varphi)}
{(1-\beta\sin\phi\sin\varphi)^2-(\beta\cos\phi\cos\varphi)^2}.
\end{equation}
This spectral property differs dramatically when an electron beam of finite size and velocity distribution is launched. The corresponding radiations now consist of two parts, the incoherent part due to direct integration of individual radiations from each electron and the coherent part due to interference at proper radiation wavelengths. The CTR is found to be much more efficient than the incoherent transition radiaiton (ITR), and its energy spectrum is strongly correlated with the beam profile as calculated by~\cite{zheng2003theoretical}:
\begin{equation}
	\label{spec_beam}
\begin{split}
\frac{d^2\xi_{CTR}}{d\omega d\Omega}=
\frac{e^2N(N-1)}{\pi^2c}|\int dtd\bm{r}d\bm{v}S(\beta, \varphi, \phi)\\
f(t, \bm{r}, \bm{v})e^{i\omega t-i\bm{k}\cdot \bm{r}}|^2,
\end{split}
\end{equation}
where $f(t,\bm{r},\bm{v})$ is a function of time, space, and velocity, describing the electron distribution within the beam. In the following, we investigate two simple models theoretically. Simulations later show that most phenomina are able to interpret by these two models.  

\subsection{CTR by a uniform electron beam}

\begin{figure}[t]
	\centering
	\includegraphics[width=0.48\textwidth]{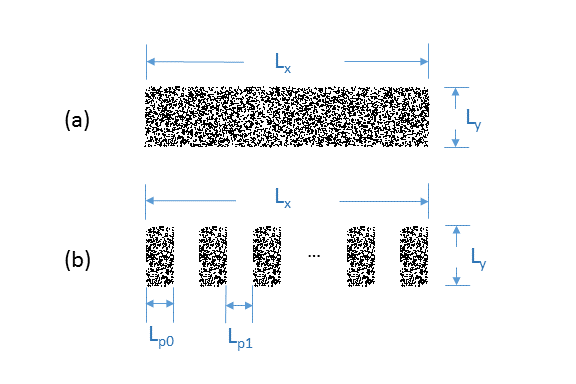}
	\caption{Geometry of (a) electron beam and (b) electron pulses train.}
	\label{electrons}
\end{figure}

In a most simple model, the electron beam has uniform density with a rectangular shape and electrons all move at the same speed with the distribution function given by $f(t, r, v)= 1/V$ for $0\leq t\leq L_x/v$ and $ |y|\leq L_y/2$.$V$ is the volume of the electrons. Geometry of the electron beam is shown in Fig.~\ref{electrons}(a).  Substitute $f(t,r,v)$ into Eq.~(\ref{spec_beam}), we get the CTR spectrum as
\begin{equation}
\label{uniformbeam_ctr}
\frac{d^2\xi_{CTR}}{d\omega d\Omega}=
 \frac{16e^2N(N-1)}{\pi^2cV^2}|S(\beta, \varphi, \phi)|^2\frac{\sin^2(\frac{kL_y}{2})}{k^2} \frac{\sin^2(\frac{\pi\omega}{\beta\omega_b})}{\omega^2}
\end{equation}
with $\omega_b = 2\pi c/L_x$. For relativistic electrons, i.e., $\beta\approx1$, TR can be coherent at frequencies $\omega<\omega_b$ or wavelengths  $\lambda>L_x$. A characteristic of TR by a single electron or ITR is that their spectra are independent of frequency. While TR by an electron beam is coherent at wavelengths longer than the electron beam size as they are emitted at roughly the same phase and can  add up coherently. From Eq.~(\ref{uniformbeam_ctr}), CTR is featured with harmonics of $\beta\omega_b$, but the intensity peaks are at $\omega=0$, about$1.43\beta\omega_{b}$ (when$\frac{\pi\omega}{\beta\omega_b}=\tan(\frac{\pi\omega}{\beta\omega_b})$) , and so on. The intensity dramatically decreases with frequency in $\omega^{-2}$, as seen in Fig.~\ref{CTR_theory}(a). $\lambda_0$ is a normalization length. $\omega_0$ is accordingly the normalization angular frequency.

\subsection{CTR by an electron beam of substructures}
\begin{figure}[t]
	\centering
	\includegraphics[width=0.35\textwidth]{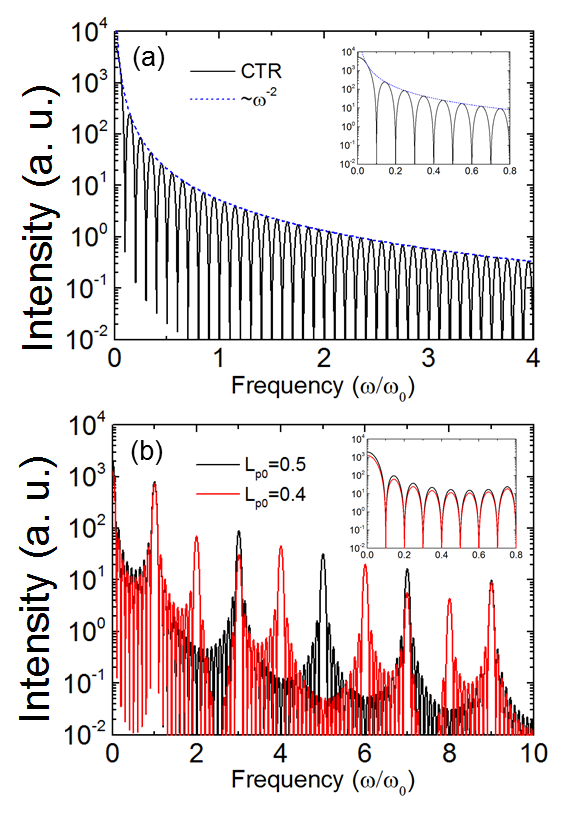}
	\caption{(Color online).Sepctra of CTR (a) from an electron beam, (b) from electron beam pulses. (a) $L_x=10\lambda_0$. Black line is the spectrum of CTR, blue dashed line marks a fit of $~\omega^{-2}$. (b) $L_p=1.0\lambda_0$, $N_p=10$.  Black and red lines are for $L_{p0}=0.5\lambda_0$ and $0.4\lambda_0$ respectively.}
	\label{CTR_theory}
\end{figure}

Electron source sometimes is a train of bunch pulses, such as the electron pulses produced by RF gun in a conventional accelerator~\cite{hoffstaetter2008compensation}, or more complicated electron sources from intense laser-plasma interactions. We consider a beam containing a train of $N_p$ identical micro-bunches of length $L_{p0}$ separated  by a  distance of $L_{p1}$, and $L_p=L_{p0}+L_{p1}$. The distribution function in this case is $f(t, r, v)= 1/V$ for $0\leq t\leq T_{p0}$, or $T_p\leq t\leq T_p+T_{p0}$, ...  or $(N_{p}-1)T_p\leq t\leq(N_{p}-1)T_p+T_{p0}$, and $ |y|\leq L_y/2$, where $T_p=L_{p}/v$, $T_{p0}=L_{p0}/v$. Substituting $f(t,r,v)$ into Eq.~\ref{spec_beam}, the CTR sepctrum from electron beam trains will be

\begin{equation}
\label{beamtrain_ctr}
\begin{split}
\frac{d^2\xi_{CTR}}{d\omega d\Omega}=
 \frac{4e^2N(N-1)}{\pi^2cV^2}|S(\beta, \varphi, \phi)|^2\frac{\sin^2(\frac{kL_y}{2})}{k^2} \frac{\sin^2(\frac{\pi\omega}{\beta\omega_{p0}})}{\omega^2} \\
 \left\{ [ \sum_{k=0}^{N_p-1}\cos(k\omega T)]^2+[\sum_{k=0}^{N_p-1}\sin(k\omega T)]^2\right\}
\end{split}
\end{equation}
The spectra are plotted in Fig.~\ref{CTR_theory}(b). There are harmonics from each pulse as well as the whole beam, i.e. harmonics of $\beta\omega_{p}$ and $\omega_b=\beta\omega_{p}/N_p$, where  $\omega_{p} = 2\pi c/L_{p}$ and $\omega_{p0} = 2\pi c/L_{p0}$. A interesting feature about CTR from elctron beam pulses is, due to the term of $\sin^2(\frac{\pi\omega}{\beta\omega_{p0}})$, when $L_{p0}=L_{p1}$, even harmonics of  $\beta\omega_{p}$ disappears, odd ones remain. A slight difference of $L_{p0}$ and $L_{p1}$ will make the even harmonics appear.
The closer of $L_{p0}$ and $L_{p1}$, the weaker the even harmonics. As in Fig.~\ref{CTR_theory}(b), when $L_{p0}=L_{p1}=0.5\lambda_0$, there is no even harmonics of $\beta\omega_{p}=\omega_0$. While $L_{p0}=0.4\lambda_0$, $L_{p1}=0.6\lambda_0$, even harmonics are produced.

\section{PIC simulations with a uniform electron beam}
In the following, we perform two-dimensional  particle-in-cell (2D PIC) simulations to investigate the transition radiation, especially coherent transition radiation, from an electron beam under varying incidence conditions. 

We start with a simplest configuration that a uniform electron beam of rectangular shape is normally incident onto an overdense plasma foil. The simulation box consists of 5000$\times$5000 cells, with 20 cells per normalization length $\lambda_0$. In this work, $\lambda_0$ can be seen as 1$\mu m$, and  the associated normalization quantities are: time $\tau_0=\lambda_0/c=3.34~\rm fs$, frequency $f_0=c/\lambda_0=300~\rm THz$ and thedensity $n_0=\epsilon_0m_e\omega_0/e^2=1.11\times10^{21}~\rm cm^{-3}$, where $\epsilon_0$ is the vacuum permittivity, $m_e$ the electron rest mass,  and $\omega_0=2\pi f_0$. If not specified, default parameters are set as follows. Electron density of the electron beam and the target is $n_b=0.001n_0$ and $n_p=10n_0$, respectively. Kinetic energy and velocity of beam electrons are $E_k=1.0~\rm GeV$ and $v=\beta c$. Length of the electron beam along the propagation direction is $L_x=5\lambda_0$, and its transverse width $L_y=5\lambda_0$. Target thickness $D=5\lambda_0$. Distance from target front where the electron beam is incident is fixed at 80$\lambda_0$. Electron charge in the beam is therefore estimated to be $44.6~\rm pC$.

\begin{figure}[t]
	\centering
	\includegraphics[width=0.48\textwidth]{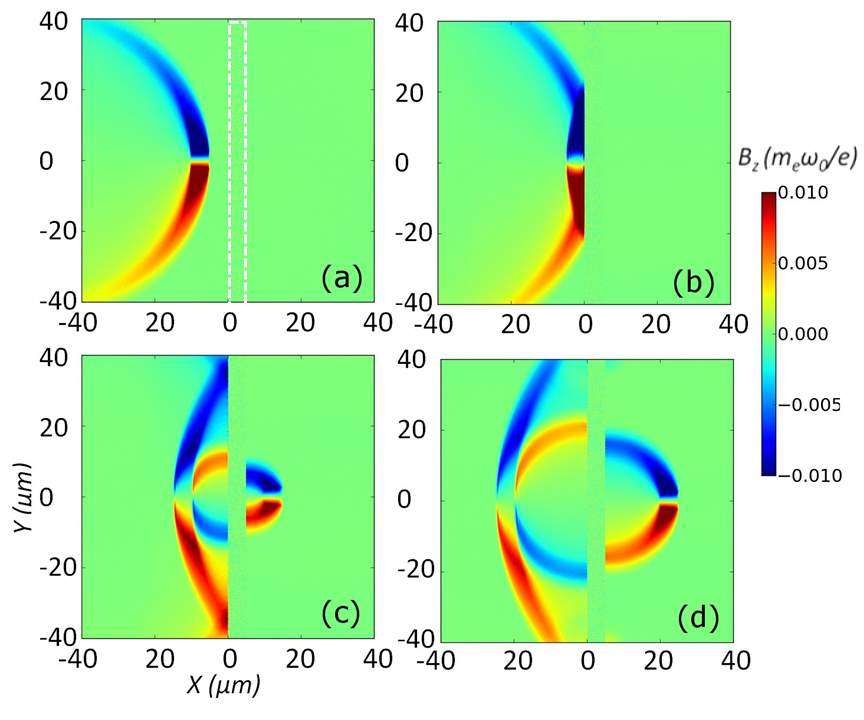}
	\caption{(Color online). Snapshots of magnetic field $B_z$ before (a), while(b) and after(c)-(d)  an electron beam hits an overdense plasma foil with its front surface placed at $x=0$. }
	\label{radiation_ebeam}
\end{figure}

Figure~\ref{radiation_ebeam} illustrates the generation of CTR by showing  the out-of-plane magnetic field $B_z$ at a few typical moments. Before the beam-foil interaction, we see two half-cycle fields of opposite polarity are associated with the cruising bunch; this is the bunch self-fields. Upon hitting the foil, strong reflections are seen, corresponding to backward CTR radiation.  The CTR shares almost the same spatial profile as the bunch self-fields except for the opposite curvature due to reflection. The backward radiation develops into a single-cycle wave imedieately. At this high beam energy, the electrons pass through the few-micrometer overdense foil almost unperturbed. During crossing the foil, the bunch self-fields are screened out by the plasma, and they only re-emerge when the bunch reaches the foil rear boundary as seen in Fig.~\ref{radiation_ebeam}(c). Interestingly, forward radiation is always half-cycle, while backward radiation is single-cycle. This is also observed in our previous study on THz radiation produced laser-solid plasma interaction~\cite{ding2013high}. 

\subsection{Angular and spectral distribution of CTR}
\begin{figure}[t]
	\centering
	\includegraphics[width=0.48\textwidth]{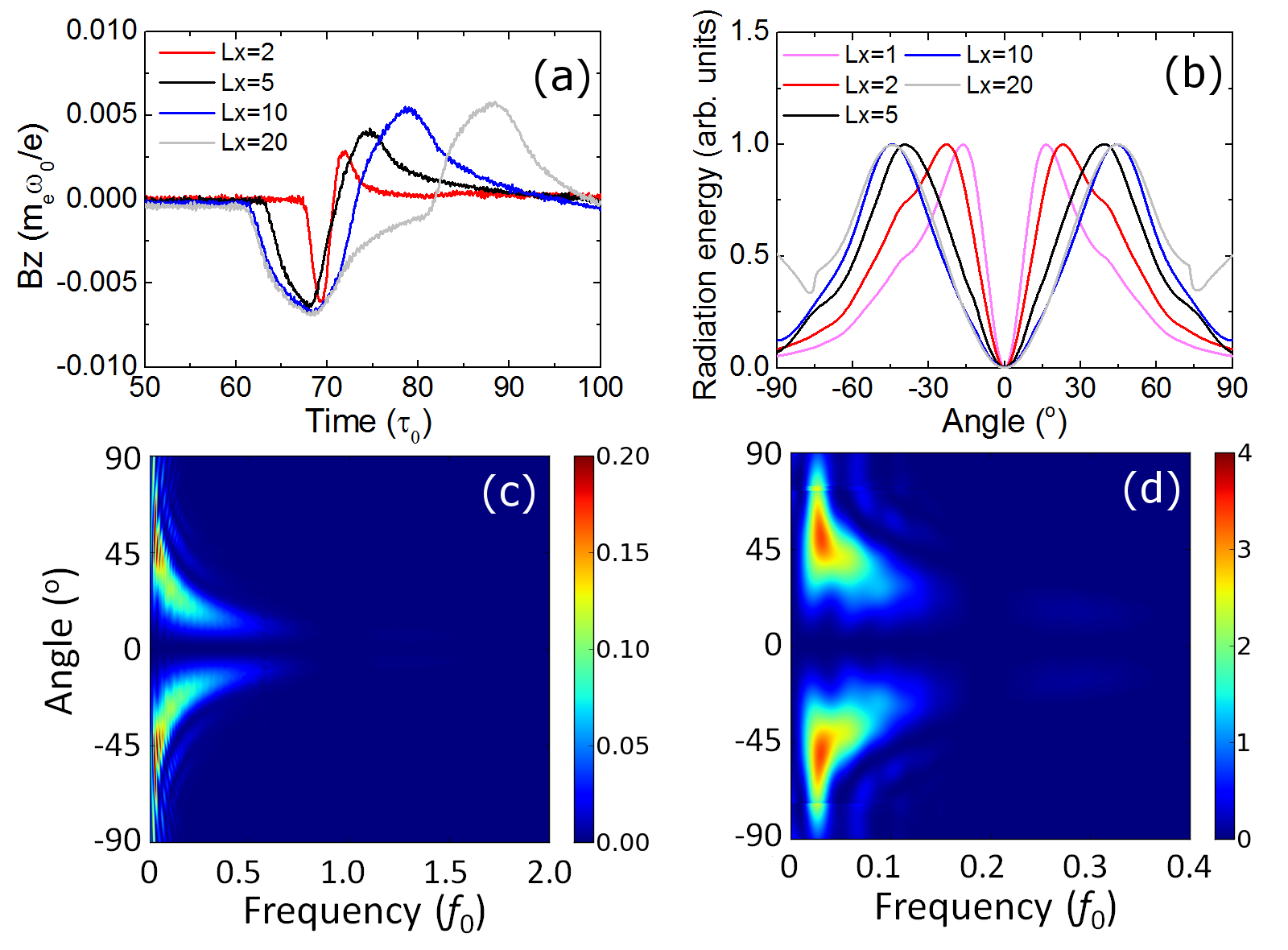}
	\caption{(Color online). (a) Temporal and (b) angular profiles of CTR for different beam lengths. The curves in (b) are rescaled to their own maxima. Angular spectra of CTR for beam length (c) $L_x=\lambda_0$ and (d) $L_x=5\lambda_0$.}
	\label{temporal_angular}
\end{figure}

Then we vary the electron beam length, $L_x$, to find out how the CTR properties change accordingly. Figure~\ref{temporal_angular} shows the temporal, angular and spectral profiles of the backward CTR obtained for different beam lengths from a fraction of to several times the foil thickness. In all cases, the reflected radiation is single-cycled with duration $2L_x/\beta c$ and coherent in the frequency from 0 to $\beta c/L_x$. The angular profiles of the radiations show a hollow cone around the electron beam path, which agrees with the theoretical predictions. However, the opening angle of the cone now varies significantly with beam length although the beam energy is ultra-relativistic. This is in sharp contrast to ITR or TR from a single electron where the radiations form a narrow cone of angle $1/\gamma$, close to the electron path under ultra-relativistic conditions. For CTR, the cone angle generally decreases for smaller $L_x$, and resembles that for single electron if $L_x$ is sufficiently small.

\begin{figure}[t]
	\centering
	\includegraphics[width=0.3\textwidth]{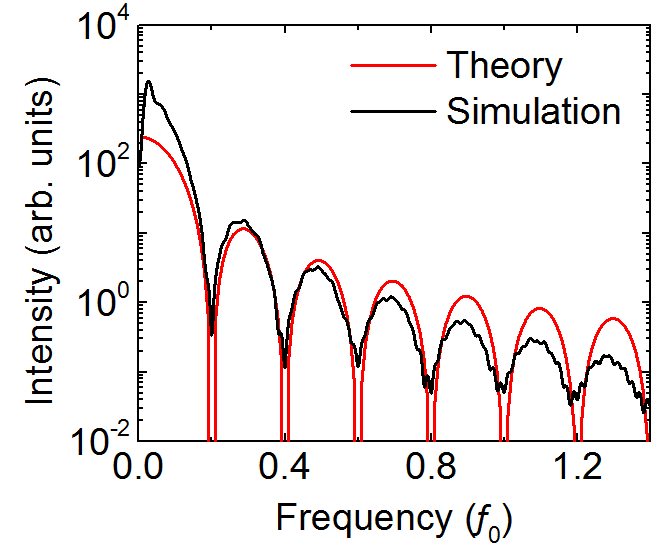}
	\caption{(Color online). Spectra of CTR from a uniform electron beam, with theory E.q.(~\ref{uniformbeam_ctr}) and simulation, respectively. $L_x=5\lambda_0$.}
	\label{spectra_uniform}
\end{figure}

CTR spectra summed on all angles are shown in Fig.~\ref{spectra_uniform}. As predicted by the analytic result E.q.(~\ref{uniformbeam_ctr}), the radiation is mainly emitted in the frequency range within $\omega_b=2\pi c/L_x$.  In the simulation, the value is as high as more than $97\%$. However, the intensity decreases  faster than the theoretical model.

According to the theory of transition radiation by a single electron, the radiation peaks at $S_{max}(\beta) = \frac{1}{(\gamma-\gamma\beta^2+1/\gamma)^2}$, when $\varphi=\sin^{-1}(1/\gamma\beta)$. It means the radiation increases rapidly with the electron energy. When $\gamma\gg 1$, $S_{max}(\beta)=\gamma^2$. However, $S(\beta, \varphi, \phi)$ with fixed $\varphi$ and $\phi$ saturates when electrons carry relativistic energy $E_{k}\gg mc^2=0.511~\rm MeV$, can be seen from Eq. (2). While CTR peaks almost at a fixed angle, which is determined by the spatial profile of the electron beam, as the electron energy increases. These predict that CTR will saturate when electron energy increases. 

\begin{figure}[t]
	\centering
	\includegraphics[width=0.48\textwidth]{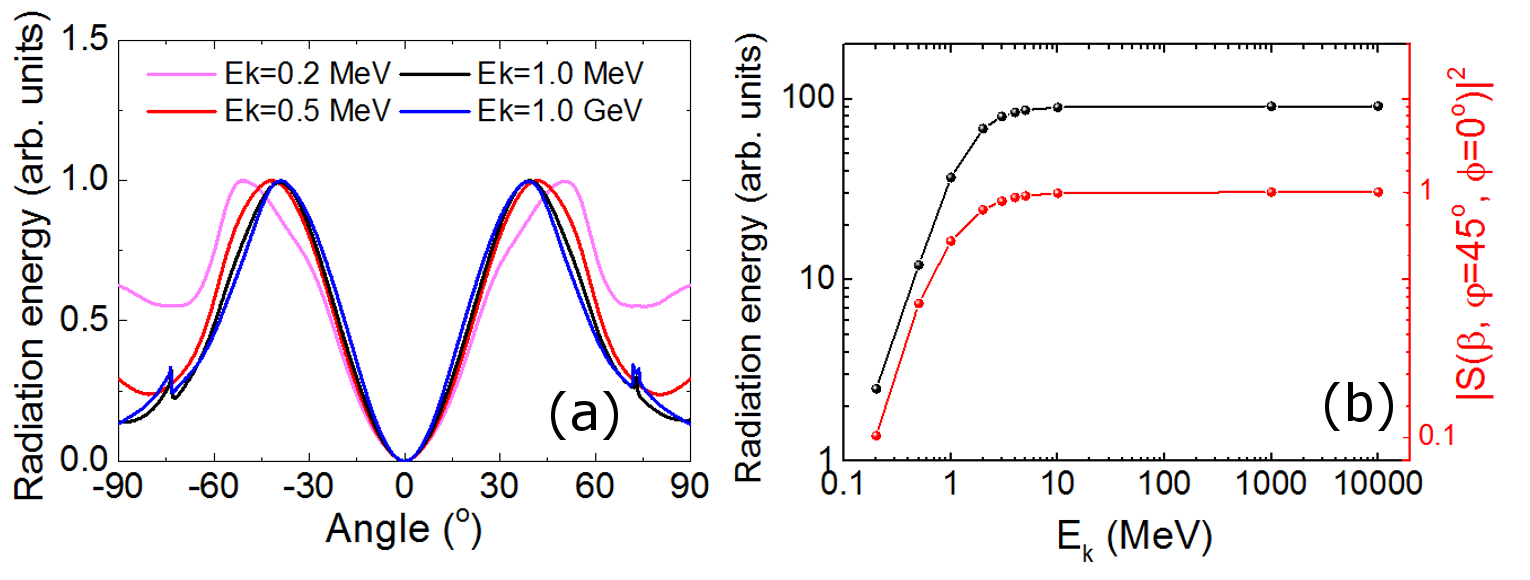}
	\caption{(Color online) (a) Angular distributions of CTR from the electron beam. Each curve is rescaled to its own maximum. (b) Radiation energy in function of electron energy. (left) Simulation results of energy of CTR from an electron beam. (right) Theoretic results of TR from a single electron at fixed direction $\varphi=45^{o}$. }
	\label{angular_energy}
\end{figure}

Simulations have reproduced the saturatioin effect as predited. Fig. ~\ref{angular_energy}(a) shows that the angular distribution of CTR does not change much as the electron energy changes. Fig. ~\ref{angular_energy}(b) shows the radiation energy saturates when electron energy $E_{k}$ is higher than a few MeV, i.e. $\gamma>10$. Radiation from a single electron at a fixed angle, for example, $|S(\beta,\varphi=45^{o},\phi=90^{o})|^{2}$ is also plotted in red line, showing the same saturation trend with CTR. 

\subsection{Saturation of CTR energy}


We have also changed the target thickness from 0.1$\lambda_0$ to 20$\lambda_0$, or the plasma density from $n_0$ to 100$n_0$, and found that the CTR has almost no change. However, later we will show that the target density has a switch-on effect on CTR generation.

\section{PIC simulations with electron pulses train}
In laser plasma interaction, multiple electron pulses are often produced, with length of one or half the laser wavelength for each pulse and total duration approximates laser duration. Similarly, we launch a train of electron beams. The thickness and spacing of the electron pulses are $L_{p0}$ and $L_{p1}$, respectively. One cycle of the micro-bunch is therefore given by $T_p= L_p/\beta c$, with $L_p= L_{p0}+L_{p1}$. The full length is still $L_x$. Simulations show that the CTR spectrum depends on these parameters of the electron beam train, especially on $L_x$ and $Lp$, same as in former theoretical model.

\subsection{Angular and spectral distribution}

\begin{figure}[t]
	\centering
	\includegraphics[width=0.48\textwidth]{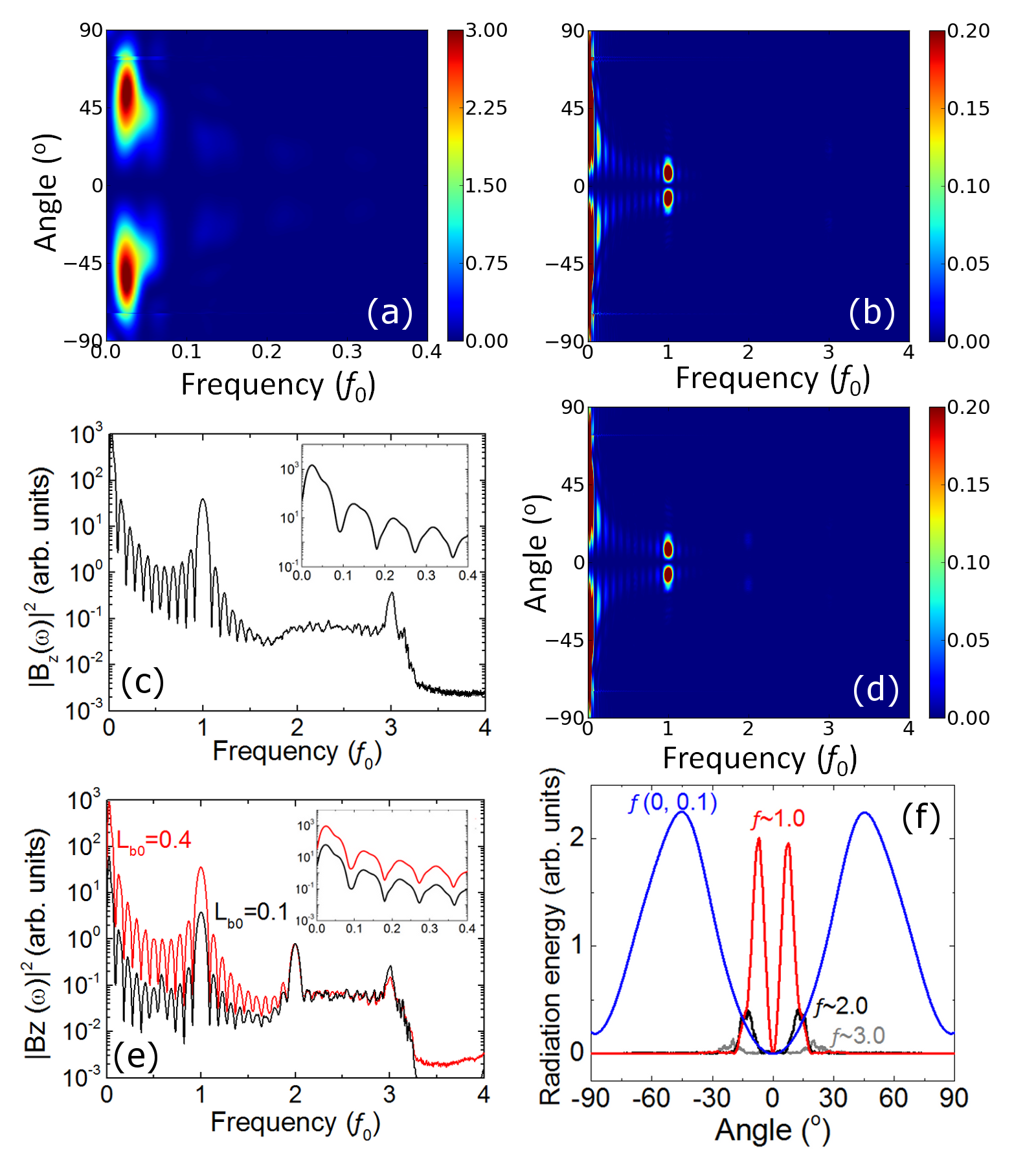}
	\caption{(Color online). Spectra of CTR from multiple electron bunches. (a) and (b) Angular distribution of the spectrum. (c) Spectrum summed on all angles, inset is a zoomed-in of the spectrum. $L_p=1.0\lambda_0$, $L_{p0}=0.5\lambda_0$, $L_x=10.0\lambda_0$. (d) Angular distribution, $L_{p0}=0.4\lambda_0$, other parameters the same. (e) Spectra of CTR with two different elctron bunches, $L_{p0}=0.4\lambda_0$ and $0.1\lambda_0$, respectively. $L_p=1.0\lambda_0$. Inset is a zoomed-in of the spectra. (f) Angular distribution of each harmonics: THz range with frequency from 0 to $0.1f_0$,  harmonics of $1.0 f_0$, $2.0 f_0$ and $3.0 f_0$. $L_{p0}=0.1\lambda_0$, $L_p=1.0\lambda_0$.
 }
	\label{harmonics_angle}
\end{figure}

Figure~\ref{harmonics_angle} shows spectra of CTR from electron pulses train. Besides harmonics of low frequency $\omega_{Lx}$ due to the whole electron train, harmonics of high frequency due to the micro structure of the electron beam is emitted as well.
Odd harmonics with frequencies of $(2m+1)\omega_p$ are produced, where $\omega_p=2\pi/T_p$, as seen in Fig.~\ref{harmonics_angle}(b) and (c). Even harmonics do not show because the length of electron pulse equals the spacing of pulses.


When the length and spacing of electron pulses are not equal, even if a smallest difference exists, e.g. $L_{p0}=0.4\lambda_0$, $L_{p1}=0.6\lambda_0$, both odd and even harmonics are produced, which are demonstrated in Fig.~\ref{harmonics_angle}(d) and (e). In Fig.~\ref{harmonics_angle}(e), the length of the electron pulse is different, e.g. $L_{p0}=0.4\lambda_0$ or $0.1\lambda_0$, while the periodic length and the total length of the electron beam train are the same, $L_p=\lambda_0$ and $L_x=10\lambda_0$.  Althought the spectrum profile is the same, raidation power with longer $L_{p0}$ is stronger, because more electrons contribute to the generation of CTR. The radiation energy by $L_{p0}=0.4\lambda_0$ is 15.3 times of that by $L_{p0}=0.1\lambda_0$ at $\omega=0.024\omega_0$, agreeing with the $N^2$ law of CTR in E.q.(~\ref{spec_beam}). 

Fig.~\ref{harmonics_angle}(f) shows that CTR with low frequency irradiates to a broad angular space, which is plotted in blue line, with summation on frequencies from 0 to $\omega_{Lx}=0.1\omega_0$ (i.e. 30THz).  Higher frequency components, $\omega_p$, $2\omega_p$ and $3\omega_p$ (i.e. 300THz, 600THz and 900THz) , are shown as well. They are calculated by summing up frequencies from $0.9\omega_0$ to $1.1\omega_0$, $1.9\omega_0$ to $2.1\omega_0$, and $2.9\omega_0$ to $3.1\omega_0$ respectively. High frequency components, due to micro structure of the elctron beam, are emitted in a narrow angular space, close to the electron beam moving direction. The difference of angular distribution for different frenquency components of CTR is consistent with the results of CTR from uniform electron beam with different length in Fig.~\ref{temporal_angular}. 

\begin{figure}[t]
	\centering
	\includegraphics[width=0.24\textwidth]{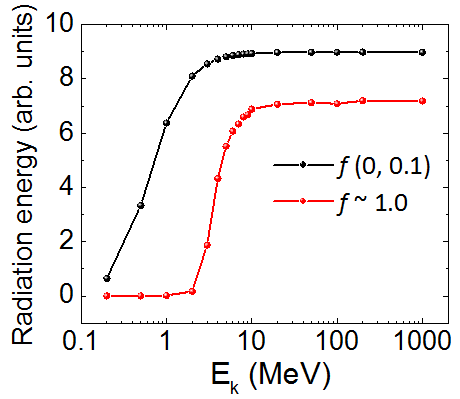}
	\caption{(Color online). CTR energy in function of plasma density. black dot line is radiation energy summed from frequency 0 to $0.1f_0$, red dot line from $0.9f_0$ to $1.1f_0$.  $L_{p0}=0.2\lambda_0$, $L_{p1}=0.8\lambda_0$, $L_x=10\lambda_0$.
}
	\label{saturation_train}
\end{figure}

We plot CTR energy with different electron energy in Fig.~\ref{saturation_train}. Similar to CTR from a uniform electron beam in Fig.~\ref{angular_energy}, CTR energy saturates when the electrons are relativistic fast. Moreover, for lower frequency of CTR, the radiation saturates at lower electron energy. For example, CTR at THz range (frequency from 0 to $0.1\omega_0$ in Fig.~\ref{saturation_train}) saturates when $E_k\geq 5~\rm MeV$, while CTR at $\lambda=\lambda_0$ saturates when $E_k\geq 10~\rm MeV$. In the laser-solid plasma interaction, electron energies are mainly within the order of $1~\rm MeV$, which is before the saturation range. Increase electron energy furtherly will help achieving higher energy THz source. As for THz radiation source via linear accelearator~\cite{DaranciangLinacCTR}, electron energy can be higher than $10 ~\rm GeV$, which is far beyond the saturation energy, and unnecessary in terms of obtaining high power THz source.

\subsection{Switch-on effect of target density}

\begin{figure}[t]
	\centering
	\includegraphics[width=0.24\textwidth]{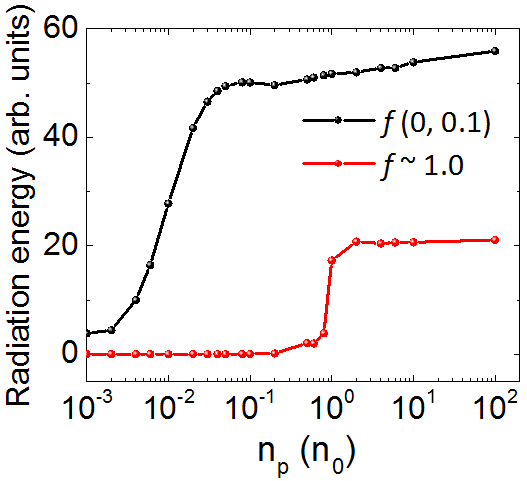}
	\caption{(Color online). CTR energy in function of target density. black dot line is radiation energy summed from frequency 0 to $0.1\omega_0$, red dot line from $0.9\omega_0$ to $1.1\omega_0$.  $L_{b0}=L_{b1}=0.5\lambda_0$, $L_x=10\lambda_0$.}
	\label{switch_density}
\end{figure}

We have mentioned in previous section that the plasma density of the target has little effect on TR, when the plasma frequency is much higher than the radiation frequency. However, if the plasma density is approaching the critical density of the radiation, the radiation will be weakened by decreasing the plasma density. It is reasonable that the radiation is emitted only when the plasma target is dense enough to support it. The switch-on effect of plasma density of the target is shown in Fig.~\ref{switch_density}. Radiation with frequency of $0.1\omega_0$ is turned on when the plasma density close to the critical density, i.e. $n_p\approx 0.01n_0$. When $n_p\gg 0.01n_0$, radiation energy saturates. Similarly, radiation with frequency of $\omega_0$ is switched on at $n_p\approx n_0$ and saturates after that. 
However, in experiments of laser- plasma interaction, different materials determine the density of both the target and the hot electrons. Therefore, radiation power changes with materials.

\section{Summary}
In summary, we have studied in detail the coherent transition radiation (CTR) from an electron beam interacting with an overdense plasma foil. Frequency and angular distribution of CTR are determined by the length of electron beam. Intensity of harmonics decreases with $1/\omega^2$, and more than $97\%$ of the energy is in the first harmonic, within the frequency of $2\pi c/L_x$. Emission angle of radiation decreases with frequency. Radiation at low frequency is emmitted to a broad angular space around the reflection direction, while higher frequency to a narrow angular space.  CTR energy saturates when electrons are highly relativistic (more than a few MeV). Increase of electron energy to GeV will not help to increase the power of CTR. CTR from electron beam pulses is investigated as well. Harmonics of $2\pi c/L_x$ and  $2\pi c/L_p$, from the whole electron beam and micro pulse, are both produced. When the length of the pulse and the spacing are close or equal, even harmonics of $2\pi c/L_p$ disappears. Otherwise, both odd and even harmonics are emitted. Target density has a threshold for the generation of CTR. It is only when the plasma frequency is equal or higher than the frequency of CTR that the CTR can be emitted. We believe our results make the generation of THz radiation much more clear, especially for the method of THz source from laser-plasma interaction. It paves the way to tunable tabletop ultra-strong THz sources. 

\section{Acknowledgements}
The work is supported by A*STAR SERC Young Individual Research Grants (YIRG No. A1784c0020).


\end{document}